# Animal behavior facilitates eco-evolutionary dynamics

The EcoEvoInteract Scientific Network

*Authors:* Gotanda, K.M.[1*] (kg419@cam.ac.uk), Farine, D.R.[2,3,4*], Kratochwil, C.F.[5,6*], Laskowski, K.L.[7*], Montiglio, P.O.[8*]

[1]Department of Zoology, University of Cambridge; www.kiyokogotanda.com @photopidge

[2]Department of Collective Behavior, Max Planck Institute of Animal Behavior, Konstanz

[3]Department of Biology, University of Konstanz

[4]Centre for the Advanced Study of Collective Behavior, University of Konstanz

[5]Chair in Zoology and Evolutionary Biology, Department of Biology, University of Konstanz

[6]Zukunftskolleg, University of Konstanz, Konstanz, Germany

[7]Department of Evolution and Ecology, University of California Davis

[8]Department of Biological Sciences, University of Quebec at Montreal

[*]All authors contributed equally

**Abstract**

The mechanisms underlying eco-evolutionary dynamics (the feedback between ecological and evolutionary processes) are often unknown. Here, we propose that classical theory from behavioral ecology can provide a greater understanding of the mechanisms underlying eco-evolutionary dynamics, and thus improve predictions about the outcomes of these dynamics.

**Eco-Evolutionary Dynamics**

The recognition that ecological and evolutionary processes can occur on the same timescale, and thus interact with each other, has led to a field of interdisciplinary research often called eco-evolutionary dynamics. Eco-evolutionary dynamics are feedbacks that occur when changes in ecological processes influence evolutionary change, which then in turn feedback onto the ecology of the system. For example, dispersal rates can increase or decrease due to ecological changes (e.g. resource fluctuations) altering species (meta-)population dynamics through source-sink dynamics or shifts in gene flow and ultimately changing the selection pressures that population experiences [1].

Eco-evolutionary dynamics have been well modeled from a theoretical perspective, drawing on the rich body of ecological literature (e.g. Lotka-Volterra models), meta-population theory, and population dynamics. However, one large body of theoretical literature is often overlooked within eco-evolutionary dynamics: behavioral ecology. Behavior lies at the heart of the decisions an individual animal makes, including if, when, and how it disperses, where it forages, how quickly it responds to potential threats, and who to mate with. Such behavioral decisions are ultimately what drives population dynamics and species interactions, making behavioral ecology a critical but perhaps under-appreciated component of eco-evolutionary dynamics.

Why should the study of eco-evolutionary dynamics more explicitly consider behavioral ecology? We argue that behavioral ecology has a robust and well-developed theoretical foundation that can provide mechanistic guidance on how systems could evolve. Models stemming from behavioral ecology—such as optimal foraging, collective decision making, kin selection, and parental care investment—can explain and predict how and why animals make the decisions that they do. Incorporating behavioral ecology models and concepts within the eco-evolutionary dynamic framework will provide mechanistic insights into how these processes are coupled and offer potentially novel alternatives for how we expect these dynamics to proceed. We demonstrate, using two examples, how incorporating behavioral ecology models can change the predictions of eco-evolutionary dynamics.

**Dispersal behavior**

The feedbacks linking dispersal to metacommunity dynamics are perhaps some of the most studied topics in eco-evolutionary dynamics [2]. Changes in ecology, such as increased habitat fragmentation, can often promote the evolution of dispersal to avoid inbreeding and/or kin competition which then increases individual fitness (Figure 1a) [3]. The evolution of an increased tendency to disperse then affects gene flow and metacommunity dynamics (e.g. population extinction probability and carrying capacity), feeding back to alter the ecology of populations, for example through changes in competition or social structure. The classic assumption in many of these eco-evolutionary models is that increases in relatedness within populations will have largely negative consequences such as inbreeding depression and mutation accumulation [4]. However, behavioral ecology can offer a novel alternative outcome to such a scenario.

One of the pillars of behavioral ecology is Hamilton's Rule, which states that cooperation can evolve if its benefits, weighted by the relatedness among cooperators, exceeds the costs of cooperating [5]. Therefore, an alternative pathway in response to habitat fragmentation is for potential dispersers to stay in their natal patch and help raise related offspring (Figure 1b). Such cooperative breeding families can pool resources allowing them to be more productive in resource poor environments and increases population fitness. The transition to cooperation is thought to be central in the colonization of novel or harsh environments, or in the persistence of populations in such environments [6]. The evolution of cooperation can then affect the relationship between the amount of resources available and population size. Whether helping evolves or not will ultimately affect the propensity for populations to persist as the habitat changes.

Behavioral ecology is underpinned by a deep understanding of how animals weigh the costs and benefits of different behaviors given the environment and their ecology. Thus, models of behavioral responses to changing environmental conditions, including the social environment (in this case kinship) provide better, or alternative, mechanistic predictions of how eco-evolutionary dynamics will proceed in a given population.

**Predator prey behavior**

Eco-evolutionary dynamics have also focused on explaining the stability, shape, and lag of predator-prey cycles [1]. Theoretical and empirical models have shown that allowing the prey and/or predator to evolve leads to changes in the synchrony of predator-prey population cycles [7]. It is generally expected that predator abundance will increase following an increase in prey abundance. In nature, however, these patterns can take different forms. These include quarter phase lags where predator abundance peaks just after peak prey abundance (Fig 2a) or antiphase lags where predator abundance peaks when prey abundance is lowest (Fig 2b) [7]. A major question is therefore to understand and ultimately predict why different predator-prey population cycles occur and how stable such cycles might be.

The stability and lag of predator-prey cycles has been linked to the shape of the predator's functional response [8]. A predator's functional response (e.g. Type II vs. Type III) is dependent on a variety of factors such as search, handling, and digestion time [9]. Such factors, and thus the functional response, will be mediated through behavior, both of the predator and importantly, of the prey. Theoretical models focusing on prey adaptations have typically considered behaviors such as activity level, refuge use, vigilance, or foraging [10,11]. We suggest that a key, additional behavioral adaptation in prey could provide an additional mechanism to understand predator-prey cycles. According to the selfish herd theory, grouping or aggregation behavior among prey can evolve in response to increased predation threats [12], yet to our knowledge, this antipredator behavior has not been explicitly considered. Grouping or aggregation of prey behavior will change factors such as search time which determine the predator's functional response. Thus, the evolution of prey grouping could impact the shape of the functional response curve of the predator, potentially affecting the shape lag and stability of predator prey cycles [8].

We here suggest that broader considerations of behavioral adaptations in prey (in this case, grouping or herding) provides additional mechanistic insight into the underlying processes shaping the functional responses. Ultimately, such considerations would deepen our understanding and improve our ability to predict the nature of predator-prey phase cycles and stability.

**Conclusion**

Here, we use two examples to highlight how explicitly considering classical behavioral ecology theory can provide viable alternative outcomes in an eco-evolutionary dynamic framework. Specifically, such theory will facilitate more robust predictions on potential outcomes of eco-evolutionary dynamics as well as provide a mechanistic basis from which to draw on. Incorporating theory from the behavioral ecology literature can help open up the mechanistic black box of eco-evolutionary dynamics to understand why animals make the decisions they do and what the consequences are. This has the potential to improve our predictions about the trajectory of eco-evolutionary feedbacks, and in some cases, as outlined above, provide novel alternative outcomes that have not been previously considered within the eco-evolutionary dynamic literature.

**References**

1       Hendry, A.P. (2017) *Eco-evolutionary Dynamics*, Princeton University Press.

2       Legrand, D. *et al.* (2017) Eco-evolutionary dynamics in fragmented landscapes. *Ecography* 40, 9–25

3       Ronce, O. (2007) How does it feel to be like a rolling stone? Ten questions about dispersal evolution. *Annu. Rev. Ecol. Evol. Syst.* 38, 231–253

4       Henry, R.C. *et al.* (2015) Dispersal asymmetries and deleterious mutations influence metapopulation persistence and range dynamics. *Evol. Ecol.* 29, 833–850

5       Hamilton, W.D. (1964) The genetical evolution of social behaviour. I. *J. Theor. Biol.* 7, 1–16

6       Cornwallis, C.K. *et al.* (2017) Cooperation facilitates the colonization of harsh environments. *Nat. Ecol. Evol.* 1, 1–10

7       Cortez, M.H. and Weitz, J.S. (2014) Coevolution can reverse predator – prey cycles. *Proc. Natl. Acad. Sci.* 111, 7486–7491

8       Daugaard, U. and Petchey, O.L. (2019) Warming can destabilize predator – prey interactions by shifting the functional response from Type III to Type II. *J. Anim. Ecol.* DOI: 10.1111/1365-2656.13053

9       Holling, C.S. (1959) The components of predation as revealed by a study of small-mammal predation of the European pine sawfly. *Can. Entomol.* 91, 293–320

10      Abrams, P.A. (2019) Foraging Behavior as a Cornerstone of Population and Community Ecology. In *Encyclopedia of Animal Behavior* 2pp. 201–208

11      Abrams, P.A. (2008) Measuring the impact of dynamic antipredator traits on predator-prey-resource interactions. *Ecology* 89, 1640–1649

12  Hamilton, W.D. (1971) Geometry for the selfish herd. *J. Theor. Biol.* 31, 295–311

**Figure 1.**

Schematic showing how considering classic behavioral theory (Hamilton's Rule) can change the outcome of predicted eco-evolutionary dynamics. Theory often predicts dispersal to increase as a result of an environmental perturbation such as habitat fragmentation to prevent inbreeding depression and mutation accumulation or to reduce competition for resources (a). By considering Hamilton's Rule, however, it will be beneficial for natal individuals to remain and cooperate instead of disperse (b). The consideration of behavioral ecology provides an alternative mechanism for whether an individual will disperse or not, affecting population and community level dynamics.

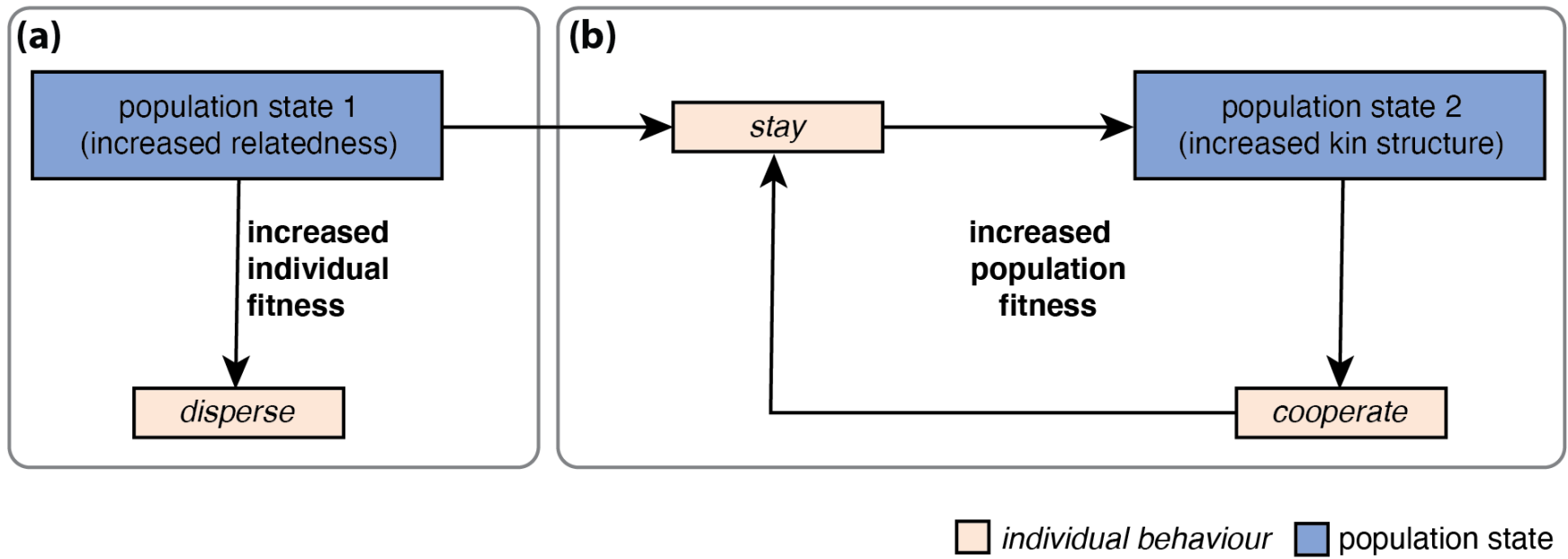

**Figure 2.**

Predator-prey cycles can have different types of lags (e.g. quarter phase (a) or antiphase (b)). The consideration of behavioral adaptations can shed light on the mechanisms underlying the dynamics and stability of predator-prey cycles.

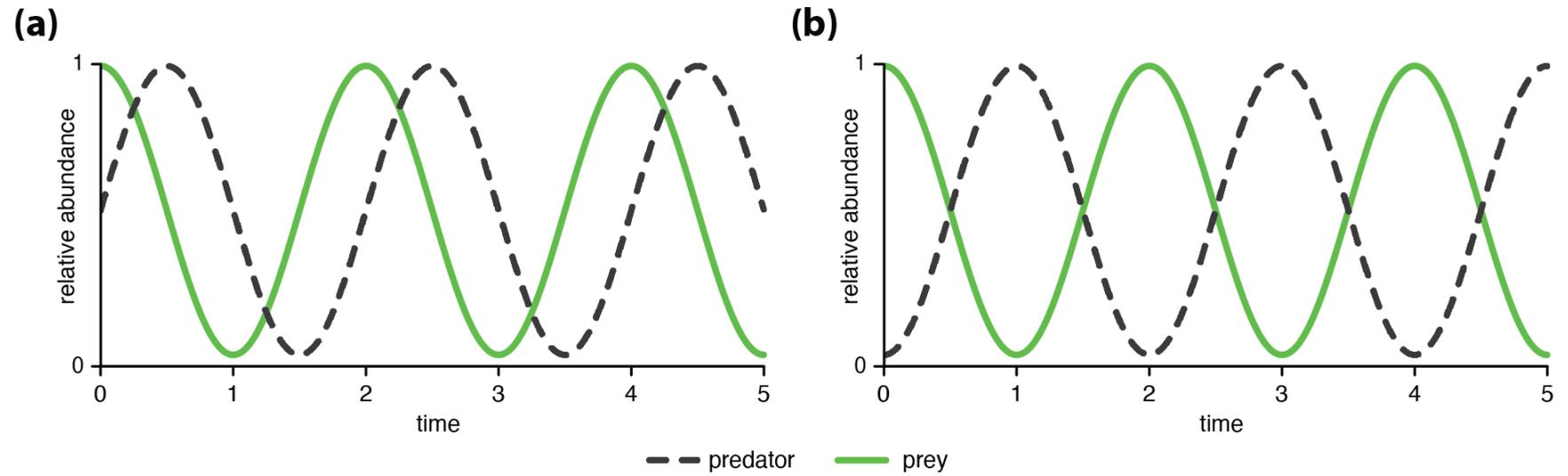